\documentclass[preprint,prd,aps,tighten,nofootinbib,amssymb]{revtex4}
\pdfoutput=1
\usepackage{graphicx}
\usepackage{epsfig,amssymb,amsmath}
\usepackage{slashed, color}
\usepackage{hyperref}
\usepackage{gensymb}
\usepackage[normalem]{ulem}
\usepackage[utf8]{inputenc}

\newcommand{\beq}{\begin{equation}}
\newcommand{\eeq}{\end{equation}}
\newcommand{\bea}{\begin{eqnarray}}
\newcommand{\eea}{\end{eqnarray}}
\newcommand{\bear}{\begin{array}}
\newcommand {\eear}{\end{array}}
\newcommand{\bef}{\begin{figure}}
\newcommand {\eef}{\end{figure}}
\newcommand{\bec}{\begin{center}}
\newcommand {\eec}{\end{center}}

\newcommand{\lsim}{\mathrel{\hbox{\rlap{\hbox{\lower4pt\hbox{$\sim$}}}\hbox{$<$}}}}
\newcommand{\gsim}{
\mathrel{\hbox{\rlap{\hbox{\lower4pt\hbox{$\sim$}}}\hbox{$>$}}}}

\begin{document}
\draft
\tighten
\preprint{CTPU-PTC-18-29}

\title{\Large \bf The dS swampland conjecture with the electroweak symmetry and QCD chiral symmetry breaking}

\author{
    Kiwoon Choi$^{a}$\footnote{Electronic address: kchoi@ibs.re.kr},
    Dongjin Chway$^{a}$\footnote{Electronic address: djchway@gmail.com},
    Chang Sub Shin$^{a}$\footnote{Electronic address: csshin@ibs.re.kr}}
     
\affiliation{
 $^a$Center for Theoretical Physics of the Universe,  Institute for Basic Science, Daejeon 34051, South Korea 
    }

\vspace{4cm}

\begin{abstract}
The dS swampland conjecture $|\nabla V|/V \geq c$, where  $c$ is presumed to be a positive constant of order unity, implies that the dark energy density of our Universe can not be a cosmological constant, but mostly the potential energy of an evolving quintessence scalar field. 
As the dark energy includes the effects of the electroweak symmetry breaking and the QCD chiral symmetry breaking,
if the dS swampland conjecture is applicable for the low energy quintessence potential, it can be applied
for the Higgs and pion potential also. On the other hand, the Higgs and pion potential has the well-known dS extrema, and applying the dS swampland conjecture to those dS extrema  may provide stringent constraints on the viable quintessence, as well as on the conjecture itself.
We examine this issue and find that
the pion dS extremum at $\cos(\pi_0/f_\pi)=-1$ 
implies $c\lesssim {\cal O}(10^{-2}-10^{-5})$ 
for  {\it arbitrary} form of the quintessence potential and couplings,
where
the weaker bound ($10^{-2}$) is available {\it only} for a specific type of quintessence whose couplings respect 
the equivalence principle, while the stronger bound ($10^{-5}$) applies for generic quintessence  violating the equivalence principle.  
We also discuss the possibility to relax this bound with an additional scalar field, e.g. a light modulus which has a runaway behavior at the pion dS extremum. We argue that such possibility is severely constrained by a variety of observational constraints which do not leave a room to significantly relax the bound.  We make a similar analysis for the Higgs dS extremum at $H=0$, which results in a weaker bound on $c$. 
\end{abstract}

\maketitle

\section{introduction} 

Motivated by the difficulty of constructing dS vacuum in string theory, recently the authors of \cite{Obied:2018sgi} proposed a conjecture that the scalar potential in low energy effective theory which has a UV completion consistent with quantum gravity
 satisfies
\bea
\label{conjecture}
M_{\rm Pl}\frac{|\nabla V|}{V} \equiv \frac{\sqrt{G^{ij}\partial_iV\partial_j V}}{V} \geq c\eea
over a certain range of scalar fields which can be of ${\cal O}(M_{\rm Pl})$,
where $G_{ij}$ is the metric of the scalar field kinetic terms in the Einstein frame, $M_{\rm Pl}\simeq 2.4\times 10^{18}$ GeV is the reduced Planck mass, and   $c$ is a positive constant of ${\cal O}(1)$. Obviously this conjecture  constrains   the possible form of (approximate) stationary points or flat directions of the scalar potential with a positive energy density. For instance, once applied for the dark energy density of the present Universe \cite{Agrawal:2018own}, it implies that the dark energy can not be a cosmological constant, but mostly the potential energy of a very light evolving  scalar field $\phi$ which is often dubbed  quintessence \cite{Ratra:1987rm,Wetterich:1987fm,Zlatev:1998tr}.

If the dS swampland conjecture is applicable for the low energy quintessence potential, it can be applied for the Higgs  and QCD pseudo-scalar meson potential also,  since the quintessence potential which is identified as the dark energy density includes the effects of the electroweak symmetry and QCD chiral symmetry breaking. 
On the other hand,  the Higgs and pseudo-scalar meson potential 
involve the well-known dS extrema, e.g. at $H=0$ or $\cos(\pi_0/f_\pi)=-1$, where $H$ is the Higgs doublet and $\pi_0$ is the neutral pion field
with the periodicity $\pi_0\equiv \pi_0+2\pi f_\pi$,
whose present vacuum values are given by $\langle H\rangle =v = 174$ GeV and $\langle \pi_0\rangle =0$. The existence of such dS extrema may impose strong constraints on the quintessence which can be compatible with the dS swampland conjecture, as well as on the parameter $c$ defining the conjecture\footnote{The dS swampland conjecture might be modified, for instance as in  \cite{Dvali:2018fqu},\cite{Andriot:2018wzk} and \cite{Garg:2018reu}, in such a way that the dS extrema that we are discussing  are manifestly compatible with the conjecture. 
}.
Indeed, it has been pointed out recently  \cite{Denef:2018etk} that if the Standard Model (SM) sector is completely decoupled from the quintessence field $\phi$, applying the dS swampland  conjecture to the Higgs extremum 
 results in $c\lesssim V(H=v)/V(H=0) \sim 10^{-55}$, which is smaller than the conjectured value $c={\cal O}(1)$   by  many orders of magnitude. 
Yet,  one can avoid this bizarre conclusion
 by assuming  proper (fine-tuned) couplings of $\phi$ 
to the Higgs sector  \cite{Denef:2018etk}, which may allow $c={\cal O}(1)$ and therefore rescue the  conjecture\footnote{For recent discussions of various implications of the dS swampland conjecture, see \cite{ds_swc}.}.

Motivated by this observation, in this paper
 we wish to examine the implications of the pion or Higgs dS extrema  for the dS swampland conjecture, while focusing
 on the possible  (model-independent or model-dependent) 
 bound on the parameter $c$. 
Here we do not question how much plausible it is to have a viable quintessence in the context of string theory, which is an issue extensively discussed  in \cite{Choi:1999xn,Choi:1999wv} a long time ago, and more recently  in \cite{Chiang:2018jdg,Cicoli:2018kdo,Akrami:2018ylq}. Instead, we take the most general quintessence potential and couplings at low energy scales,
and examine what would be the maximal value of the parameter $c$ allowed by the observational constraints. 
 We then find that the most stringent bound comes from the pion extremum at $\cos(\pi_0/f_\pi)=-1$, yielding 
\bea
\label{bound_on_c}
c \,\leq\,  {\rm Max}\left[\, d_{q} + 3 d_g ,\,  V_{\rm eff}/f_\pi^2 m_\pi^2\sim 10^{-43}\right],
\eea 
where $d_q$ and $d_g$ are the low energy quintessence couplings to the light quarks and gluons defined in (\ref{low_coupling}), $V_{\rm eff}\sim (2\times 10^{-3}\,{\rm eV})^4$ is the quintessence potential energy 
  in the present Universe, and $m_\pi$ and $f_\pi$ are the pion mass and decay constant, respectively. The observational bounds on the quintessence couplings $d_q$ and $d_g$ depend on whether they respect or violate the equivalence principle. For a specific type of quintessence whose couplings respect the equivalence principle, e.g. a quintessence  which
    couples to the SM sector {\it only} through the  trace of the energy momentum tensor, we have $d_q=d_g$.
    In such case, the quintessence couplings are constrained mainly by
    the observational bound on the deviation
  from the general relativity  by the quintessence-mediated force in relativistic limit \cite{Bertotti:2003rm}, which results in\footnote{All experimental bounds used in this paper are the 95\% confidence level bounds.}  
  \bea
  c  < 1.4\times 10^{-2} 
  \eea
  for  {\it arbitrary} form of the quintessence potential.
 We call such quintessence  a {\it metrical quintessence} \cite{Choi:1999wv} as such a specific form of couplings which respect the equivalence principle  may arise through the mixing with the conformal factor of the spacetime metric.
However, for more generic quintessence with $(d_g-d_q)/(d_g+d_q)={\cal O}(1)$,  the quintessence couplings are bounded by the non-observation of the violation of the equivalence principle in non-relativistic limit \cite{Touboul:2017grn, Berge:2017ovy}, yielding a much stronger bound \bea
c <  2\times 10^{-5}\eea
 again for {\it arbitrary} form of the quintessence potential.  The bounds on $c$ from the Higgs extremum is  weaker than those from the pion extremum, but yet significantly stronger than the results obtained  in \cite{Denef:2018etk}.

In fact, the above bounds on $c$ are obtained while assuming that other scalar fields in the underlying theory can be  integrated out without affecting the low energy dynamics around the pion dS extremum.  One can then contemplate the possibility that those bounds are relaxed by an additional scalar field, e.g. a light modulus-like scalar, which has a large tadpole or a runaway behavior when the pion field is at the dS extremum. We examine this possibility also, and find
that such a light scalar is severely constrained by a variety of 
observational constraints which practically close the room to significantly relax the above bounds on $c$.
 
Our bounds on $c$ from the pion extremum appear to have a significant tension with the dS swampland conjecture (\ref{conjecture}) which assumes $c={\cal O}(1)$. We note that the conjecture (\ref{conjecture}) can be modified or refined, for instance as in  \cite{Dvali:2018fqu},\cite{Andriot:2018wzk} and \cite{Garg:2018reu}, in such a way to avoid the bounds from the Higgs and pion extrema. Then our results can be interpreted as providing additional motivation for such refinement of the conjecture.

The organization of this paper is as follows. In the next section, we discuss the possible couplings of quintessence to the SM sector and summarize the relevant observational constraints on the quintessence couplings. In Sec. \ref{ds_extrema}, we apply the dS swampland conjecture for the pion and Higgs extrema, and examine what would be the maximal value of $c$ allowed by the observational constraints in the context of the most general form of the low energy quintessence potential and couplings.   In Sec. \ref{tadpole}, we examine if an additional scalar which has a sizable tadpole or runaway behavior at the pion dS extremum can relax the bound on $c$ obtained in 
Sec. \ref{ds_extrema}.
     Sec. \ref{conclusion} is the conclusion.

\section{\label{q_couplings}quintessence couplings to the standard model}

In this section we briefly discuss the possible couplings of the quintessence scalar field $\phi$ to the SM sector, as well as the observational constraints  on the couplings. Without loss of generality, using appropriate field redefinitions, one can always move to the Einstein frame and make
the kinetic terms of the SM  fermions and the Higgs boson to take the {\it $\phi$-independent} canonical form. We are  interested in the possible {\it non-derivative} couplings of $\phi$ to the SM fields in this field basis, which can be encoded in the $\phi$-dependent SM parameters.  Then the Lagrangian density at some scale $\mu$  above the weak scale can be written as\footnote{For simplicity, here we do not consider the quintessence couplings such as
$\theta(\phi)F^{a\mu\nu}\tilde F^a_{\mu\nu}$,   $\partial_\mu\phi H^\dagger D^\mu H$ and $\partial_\mu\phi\bar\psi\gamma^\mu\psi$ 
as they do not affect our subsequent discussion. Note that the derivative couplings of $\phi$ to $H$ and $\psi$ can affect the scalar field metric $G_{ij}$ that appears in the dS swampland conjecture, but their effects are suppressed by $v/M_{\rm Pl}\sim 10^{-16}$ or $f_\pi/M_{\rm Pl}\sim 10^{-19}$ and therefore can be safely ignored.}  
\bea\label{eq:Lagrangian}
{\cal L}&=&\frac{1}{2}\partial_\mu\phi\partial^\mu\phi 
+ |D_\mu H|^2 + \bar\psi_L iD\hskip -0.28cm\slash\psi_L
+  \bar\psi_R iD\hskip -0.28cm\slash\psi_R -\frac{1}{4g_a^2(\phi)}F^{a\mu\nu}F^a_{\mu\nu}
\nonumber\\
&&- (y_\psi(\phi) H\bar\psi_L\psi_R+ h.c.)  
- \lambda(\phi)|H|^4 +m_H^2(\phi)|H|^2 -V_b(\phi),
\eea
where $V_b$ is the $H$-independent bare potential of $\phi$, and 
$g_a(\phi)$, $\lambda(\phi)$ and $y_\psi(\phi)$ are generically  $\phi$-dependent gauge,  Higgs quartic, and Yukawa couplings,  respectively. 

From the above Lagrangian density, 
one can calculate the low energy consequences of the model, including the effective potential of $\phi$ at cosmic scales 
and also the low energy couplings of $\phi$  
 which are constrained by a variety of laboratory,  astrophysical and cosmological observations  \cite{Godun:2014naa,Touboul:2017grn,Bertotti:2003rm,Uzan:2010pm}.
For instance, the low energy quintessence potential can be obtained by integrating out all SM fields, which would result in 
\bea \label{eq:dark_energy}
V_{\rm eff}(\phi) &=& V_b(\phi) + \left\langle \lambda |H|^4 -m_H^2|H|^2
 -\left(y_\psi H\bar\psi_L\psi_R+h.c\right) +\frac{1}{4g_a^2} F^{a\mu\nu}F^a_{\mu\nu} \right\rangle+ ...
 \nonumber \\
 &=& V_b(\phi) -\frac{m_H^4(\phi)}{4 \lambda(\phi)} -\sum_{q=u,d,s}m_q(\phi)\langle \bar q q\rangle +{\cal O}(\Lambda^4_{\rm QCD})+ ...,
\eea
where $\langle ..\rangle$ are the expectation values of the SM fields,  ${\cal O}(\Lambda_{\rm QCD}^4)$ denotes the contribution from the gluon condensation including the contributions from the heavy quark thresholds effects, and
the ellipsis stands for additional contributions including a variety of additional quantum corrections.
If $\phi$ is identified as the quintessence scalar field explaining the dark energy of the present Universe, its low energy potential should satisfy \cite{Agrawal:2018own,Heisenberg:2018yae}
\bea
\label{quint1}
V_{\rm eff}(\phi) \sim (2\times 10^{-3}{\rm eV})^4,  \quad  \quad   M_{\rm Pl}\frac{V^\prime_{\rm eff}(\phi)}{V_{\rm eff}(\phi)}
 \lesssim 0.6\eea
 over a field range  $\Delta \phi \sim M_{\rm Pl}$, 
 where the prime denotes the derivative w.r.t $\phi$, 
 and also
\bea
\label{quint2}
 V^{\prime\prime}_{\rm eff}(\phi_0)  \lesssim  H_0^2 \sim (10^{-33} \,{\rm eV})^2,\eea
 where $\phi_0$ and $H_0$ are the quintessence field value and the Hubble expansion rate of the present Universe, respectively.

As for the couplings of $\phi$ defined at high energy scale $\mu$,  one finds
\bea
\label{high_coupling}
{\cal L}_{\phi}(\mu) =\frac{g_a^\prime}{2g_a^3} \phi F^{a\mu\nu}F^a_{\mu\nu} 
-\left(m_\psi\left(\frac{y_\psi^\prime}{y_\psi}+\frac{v^\prime}{v}\right) \phi\bar\psi_L\psi_R
+h.c\right)  
- m_h^2\left(\frac{\lambda^\prime}{\lambda}+\frac{2v^\prime}{v}\right)\phi h^2 + ... ,
\eea
where $v$ denotes the $\phi$-dependent Higgs vacuum value given by \bea
v^2(\phi)=\frac{m_H^2(\phi)}{2\lambda(\phi)},\eea $h$ is the canonically normalized  Higgs boson fluctuation,  and again the prime denotes the derivative w.r.t $\phi$.  Here the field $\phi$ corresponds to the fluctuation around $\phi_0$, and  
the ellipsis stands for additional couplings which are not relevant for our subsequent discussion.

As they are even weaker than the gravitational coupling,  the quintessence couplings are constrained mostly by the macroscopic observations such as the violation of the equivalence principle in non-relativistic limit,
the deviation from the general relativity in relativistic limit, or the variation of the fundamental constants, e.g. the fine structure constant \cite{Godun:2014naa}. However, those constraints apply for the low energy effective couplings of $\phi$
defined at lower energy scale $\mu_{\rm eff}$, which may be parametrized  as
\bea
\label{low_coupling}
{\cal L}_{\phi}(\mu_{\rm eff})\,=\, d_\gamma \frac{\phi}{M_{\rm Pl}} \frac{1}{4 e^2}F^{\mu\nu}F_{\mu\nu} - d_g\frac{\phi}{M_{\rm Pl}} \left(T^\mu_\mu\right)_{\rm QCD} - \sum_{q=u,d,s} (d_q-d_g)m_q \frac{\phi}{M_{\rm Pl}} \bar q q\nonumber \\
\,=\,
d_\gamma \frac{\phi}{M_{\rm Pl}} \frac{1}{4 e^2}F^{\mu\nu}F_{\mu\nu} - d_g  \frac{\phi}{M_{\rm Pl}} \frac{\beta_s}{2g_s }G^{i\mu\nu}G^i_{\mu\nu}
- \sum_{q=u,d,s}(d_q +\gamma_m d_g) m_q\frac{\phi}{M_{\rm Pl}} \bar q q,
\eea
where
$(T^\mu_\mu)_{\rm QCD}$ is the trace of the energy momentum tensor for the low energy QCD of the light quark flavors $q=(u,d,s)$, and therefore $\beta_s$ and $\gamma_m$ are the QCD beta function and the mass anomalous dimension, respectively.
Here the gluon fields $G^i_{\mu\nu}$  are rescaled to have the standard canonical kinetic term
 as $-\frac{1}{4} G^{i \mu \nu} G^i_{\mu \nu}$.
Note that $d_g$ and $d_q$ are defined to be independent of the  renormalization scale $\mu_{\rm eff}$.  Also, they can be defined 
as 
\bea
\frac{d_g}{M_{\rm Pl}}= \frac{\Lambda^\prime_{\rm QCD}(\phi)}{\Lambda_{\rm QCD}(\phi)},\qquad  \frac{d_q}{M_{\rm Pl}} =\frac{m_q^\prime(\phi,\Lambda_{\rm QCD}(\phi))}{m_q(\phi, \Lambda_{\rm QCD}(\phi))},\eea 
where $\Lambda_{\rm QCD}(\phi)$ is the $\phi$-dependent physical QCD scale and $m_q(\phi,\Lambda_{\rm QCD}(\phi))$
is the $\phi$-dependent light quark mass renormalized at $\Lambda_{\rm QCD}$.

Although the low energy couplings $d_g$ and $d_q$ are defined to be independent of $\mu_{\rm eff}$, perturbative calculation of those couplings in terms of the high energy couplings  in (\ref{high_coupling}) can be done only for  $\mu_{\rm eff}$ where the perturbation theory applies.  For later use, let us briefly discuss the perturbative corrections that $d_g$ receives from the high energy couplings in (\ref{high_coupling}).
At one-loop order, the dominant corrections to $d_g$ come from the one-loop thresholds of heavy quarks which couple to
$\phi$.  There can be also a potentially important two-loop  correction induced by
the $\phi-h-h$ coupling in (\ref{high_coupling}) and the top quark Yukawa coupling of the Higgs boson $h$. Putting those radiative corrections with the tree level contribution, we find that the low energy coupling $d_g$  
at $\mu_{\rm eff}$ just below the charm quark mass is determined by the high energy couplings in (\ref{high_coupling}) as follows:
\bea\label{eq:d_g}
\frac{d_g}{M_{\rm Pl}} \simeq    \frac{(16\pi^2)}{9}\frac{g_s^\prime(\mu)}{g_s^3(\mu) } +  \frac{2}{27} \sum_{q=t,b,c} \left( \frac{y_q'}{y_q}+ \frac{v'}{v} \right)  - \frac{ y_t^2 f( \tau) }{288\pi^2}  \left( \frac{\lambda'}{\lambda}+ 2\frac{v'}{v} \right), 
\eea 
where $\tau \equiv 4m_t^2/m_h^2$.
Here the second term in the RHS represents the one-loop threshold  of heavy quarks, while the third term corresponds to the two-loop threshold involving the Higgs boson and the top quark.
We obtained the analytic form of the function $f(\tau)$ for a generic $\tau$ from full two-loop calculations. For $m_t=173$ GeV and $m_h=125$ GeV, it yields $f(4m_t^2/m_h^2)= 0.21$.

%
%

For generic low energy quintessence couplings,
the most stringent constraint  comes from the violation of the weak equivalence principle (EPV)
by the quintessence-mediated force in non-relativistic limit. 
For instance, using the results of \cite{Berge:2017ovy,Damour:2010rm,Damour:2010rp}, we find that non-observation of EPV implies
\bea
\label{epv_bound}
(d_g+0.093(d_{\tilde{q}}-d_g)+0.00027d_\gamma) \left( 3.3(d_{\tilde{q}}-d_g) + 1.9 d_\gamma \right) < 2.7 \times 10^{-11},\eea
where \bea
d_{\tilde{q}} = \frac{m_u d_u + m_d d_d}{m_u+m_d}.\eea 
If we assume that there is no significant cancellation among the different quintessence couplings, e.g.
\bea
\frac{d_\alpha-d_\beta}{d_\alpha+d_\beta}={\cal O}(1) \quad (\alpha,\beta=g, q, \gamma)\eea
which would be the case for generic forms of low energy couplings, 
this implies
\bea \label{eq:dg_bound}
d_{g}< 3\times 10^{-6}, \quad d_{\tilde q} < 10^{-5}, \quad d_\gamma < 2\times 10^{-4}.\eea
The quintessence coupling to the photon can be constrained by
the observational bound on the time-varying fine structure constant also \cite{Godun:2014naa}, which would result in
\bea
\label{fine_bound}
d_\gamma  <3 \times 10^{-7} \frac{M_{\rm Pl}H_0}{\dot \phi},\eea
where $H_0$ is the Hubble expansion rate today  and $\dot\phi=d\phi/dt$.

In fact, there is a specific type of quintessence which automatically satisfies the above constraints from EPV and time-varying fine structure constant.
If $\phi$ couples to the SM {\it only} through the trace of the energy momentum tensor, i.e.
\bea
\label{metrical}
{\cal L}_\phi = d_T \frac{\phi}{M_{\rm Pl}} T^\mu_\mu,\eea
we have
\bea
d_T =d_g=d_q,\quad d_\gamma=0,
\eea therefore the observational bounds (\ref{epv_bound}) and (\ref{fine_bound}) are automatically satisfied.  
Note that the quintessence coupling
to $T^\mu_\mu$ does not violate the equivalence principle, and also the time-varying fine structure constant applies for the low energy electromagnetic coupling which has a vanishing beta function, and therefore the corresponding $d_\gamma=0$ when 
the quintessence couplings take the form (\ref{metrical}).

One may call the above type of quintessence a ``metrical quintessence"  since 
the specific coupling (\ref{metrical}) can arise from the mixing of $\phi$
with the conformal factor of the spacetime metric $g_{\mu\nu}$ \cite{Choi:1999wv}. A specific such example is  a theory  which does {\it not} have any coupling between $\phi$ and the SM fields in an appropriate field basis, while having non-trivial couplings between $\phi$ and $g_{\mu\nu}$ through the $\phi$-dependent Planck mass. One can then move to the Einstein frame by making an appropriate Weyl transformation:
\bea
g_{\mu\nu}  \rightarrow \Omega(\phi) g_{\mu\nu}\eea
which would result in the quintessence coupling (\ref{metrical}) in the Einstein frame.
Yet, the coupling of metrical quintessence is constrained by the observational bounds on the deviation from the general relativity by the quintessence-mediated force in relativistic limit. For instance, from the measurement of the gravitational time delay effect to the Cassini spacecraft, one finds \cite{Bertotti:2003rm}
\bea 
\label{gr_bound}
d_T=d_g=d_q < 3.4\times 10^{-3}.
\eea


\section{\label{ds_extrema} de Sitter swampland conjecture for the pion and Higgs extrema}

As we have stressed, if the dS swampland conjecture (\ref{conjecture}) applies for the low energy quintessence potential
(\ref{eq:dark_energy}) including the contributions from the electroweak symmetry and QCD chiral symmetry breaking, it is applicable also for the Higgs  and QCD pseudo-scalar meson potentials. 
In this section,
we apply the dS swampland conjecture to some of the dS extrema of the pseudo-scalar meson or Higgs potential and examine 
its implications. For simplicity, we will take the simple effective field theory approach, assuming that 
all  other degrees of  freedom can be integrated out in such a way that the resulting  effective theory is  good enough
over a field range  including both the vacuum configuration and the relevant dS extrema, e.g. 
the entire field range of the pion field $\pi_0/f_\pi \in [0,2\pi]$ and also the Higgs field range
$\Delta H \sim v$.

 Because we are considering both the vacuum solution and a dS extremum together,
  generically our results can receive corrections from the tadpoles or runaway behavior  
 of the integrated scalar fields, which can be induced at the dS extremum point.
We will see  in the next section that those corrections do not significantly affect the results of this section when the observational constraints on the underlying dynamics are properly taken into account.

\subsection{Pion extremum}

To proceed, let us first consider the field configuration where the Higgs field is frozen at its vacuum value, $H=v(\phi)=m_H(\phi) / \sqrt{2\lambda(\phi)}$, and integrate out all SM fields heavier than the QCD scale $\Lambda_{\rm QCD}$.
The remained light scalar degrees of freedom are the quintessence field $\phi$ and the pseudo-scalar meson
octet $\pi_a=(\pi, K, \eta)$ which correspond to the pseudo-Nambu-Goldstone bosons associated with the  spontaneous breakdown of the  QCD chiral symmetry:
\bea
SU(3)_L\times SU(3)_R\rightarrow SU(3)_V.\eea
 The corresponding field manifold  $SU(3)_L\times SU(3)_R/SU(3)_V$ is compact and can be parametrized  as
\bea 
U = \exp \left[ i\frac{\pi_a}{f_\pi}\lambda_a \right],
\eea 
where $\lambda_a$ are the Gell-Mann matrices and $f_\pi$ is the pion decay constant.
At leading order in chiral perturbation theory, the effective Lagrangian of $U$ is given by
\bea 
\frac{f_\pi^2}{4} {\rm Tr}\left[  \partial_\mu U^\dagger \partial^\mu U \right] + \frac{\Lambda^3}{2}  {\rm Tr}\left[ 
M_q (U+U^\dagger)\right],   
\eea 
where $\Lambda$ can be identified as the condensation scale of the light quark fields, i.e. $\langle \bar q_i q_j\rangle =\Lambda^3\delta_{ij}$ for $q_i=(u,d,s)$, and
$M_q={\rm diag}(m_u, m_d, m_s)$ is the light quark mass matrix which is chosen to be real, diagonal and positive.

Because the meson field manifold $SU(3)_L\times SU(3)_R/SU(3)_V$
is compact, there can be multiple dS extrema of the meson potential. Here, for simplicity we focus on the neutral pion 
$\pi_0$, while fixing all other mesons at their vacuum values.
Then the effective potential of the pion and quintessence is given by
\bea \label{eq:quint_pion}
V(\phi, \pi_0/f_\pi) = 
V_{\rm eff}(\phi) + (m_u(\phi)+ m_d(\phi))\Lambda^3(\phi)\left[1 -  \cos\left(\frac{\pi_0}{f_\pi(\phi)}\right)\right], 
\eea 
where we choose the convention that $\langle \pi_0\rangle=0$ in the true vacuum, and 
the low energy QCD parameters $m_{u,d}$, $f_\pi$ and $\Lambda$ are understood to be generic functions of the quintessence field $\phi$.
This potential is valid over the full range of the pion field $\pi_0/f_\pi \in [0, 2\pi]$ and has a dS local maximum along the pion direction at
\bea
\label{dS_pi}
\frac{\pi_0}{f_\pi} = \pi.
\eea
Note that although we consider a leading order approximation in chiral perturbation theory, the periodicity of the pion field
$\pi_0\equiv \pi_0+2\pi f_\pi$ and the CP invariance under $\pi_0\rightarrow -\pi_0$  assure that this configuration is a dS local maximum of the exact pion potential up to negligible corrections  due to the CP violating weak interactions.
We then find 
\bea 
V(\phi, {\pi_0}/{f_\pi}=\pi) &=& V_{\rm eff}(\phi) + 2(m_u(\phi) +m_d(\phi))\Lambda^3(\phi)\nonumber \\
\nabla V(\phi, {\pi_0}/{f_\pi}=\pi) & = & V'_{\rm eff}(\phi) 
+ 2\left(\frac{m_u'+m_d'}{m_u+m_d} + 3\frac{\Lambda'}{\Lambda}\right)(m_u+m_d)\Lambda^3,\eea
which results in
\bea
\label{eq:pion_ds_bound}
M_{\rm Pl}\frac{|\nabla V(\phi,{\pi_0}/{f_\pi}=\pi) |}{V(\phi, \pi_0/f_\pi=\pi)} \,= \,\left|M_{\rm Pl}\left(\frac{m_u' + m_d'}{m_u + m_d} + \frac{3\Lambda'}{\Lambda}\right)+ {\cal O}(10^{-43})\right|\,\geq\, c,
\eea
where we used the properties (\ref{quint1}) of $V_{\rm eff}(\phi)$ yielding
\bea
M_{\rm Pl}\frac{V_{\rm eff}^\prime}{(m_u+m_d)\Lambda^3}\,\lesssim\, \frac{V_{\rm eff}}{(m_u+m_d)\Lambda^3}\,=\, \frac{V_{\rm eff}}{m_\pi^2 f_\pi^2}\,\sim\, 10^{-43},
\eea
and applied the dS swampland conjecture (\ref{conjecture}) in the last step.

For us, it is most convenient to choose the renormalization scale of the light quark mass $m_q$ and the quark bilinear operator $\bar qq$ to be $\Lambda_{\rm QCD}$, for which
\bea 
\frac{\Lambda'}{\Lambda}=  \frac{\Lambda^\prime_{\rm QCD}}{\Lambda_{\rm QCD} }=\frac{d_g}{M_{\rm Pl}},\qquad 
\frac{m_q^\prime}{m_q} =\frac{d_q}{M_{\rm Pl}},
\eea 
where $d_g$ and $d_q$ are the low energy quintessence couplings defined in (\ref{low_coupling}).
Then the dS swampland conjecture applied for the pion and quintessence potential at the pion dS extremum results in 
\bea
\label{bound_pion_ex}
c \,\lesssim\,  {\rm Max}\left[\, d_{\tilde q} + 3 d_g ,\, {\cal O}(10^{-43})\right],
\eea 
where
\bea
d_{\tilde{q}} = \frac{m_u d_u + m_d d_d}{m_u+m_d}.\eea 
We stress that the above bound is valid for {\it arbitrary} form of the quintessence potential and couplings.

Similarly to the case of the Higgs extremum discussed in \cite{Denef:2018etk}, if $\phi$ is completely decoupled from the QCD sector, so that $d_q=d_g=0$, the parameter $c$ is required to be smaller than $V_{\rm eff}/f_\pi^2 m_\pi^2\sim 10^{-43}$, which is smaller than the conjectured value $c={\cal O}(1)$ by  many orders of magnitude. 
Again, by assuming appropriate form of couplings between 
the quintessence and the QCD sector, one can alleviate this bound on $c$ up to the value allowed by observational constraints.
Then, for generic quintessence with $(d_q-d_g)/(d_q+d_g)={\cal O}(1)$, the observational bound (\ref{epv_bound})
on the violation of the equivalence principle (EP) can be applied to get
\bea\label{eq:bound_gen}
c \,<\, 2\times 10^{-5} \quad \mbox{for quintessence violating the EP}.\eea
On the other hand,
for a metrical quintessence which couples to the SM only through the trace of energy momentum tensor and therefore has $d_g=d_q$,
the bound on $c$ can be significantly relaxed.  In such case, we can use the observational bound (\ref{gr_bound}) 
on the deviation from the general relativity in the solar system to get
\bea\label{eq:bound_met}
c \,<\, 1.4\times 10^{-2} \quad \mbox{for quintessence respecting the EP}.\eea


\subsection{Higgs extremum}

Let us now consider the Higgs dS extremum at $H=0$, which was discussed also in \cite{Denef:2018etk}. Here we will elaborate the discussion of \cite{Denef:2018etk} and examine if any useful bound on $c$ can be obtained  from the consideration of the Higgs dS extremum. 
In the scalar field space near $H=0$,  the effective potential can be written as 
\bea 
V(H, \phi)= V_{\rm eff}(\phi)+ \lambda(\phi) v^4(\phi) 
   -\sum_\psi\left( y_\psi(\phi) H\langle \bar\psi_L\psi_R\rangle +h.c\right)- \frac{1}{2}m_H^2(\phi) |H|^2 +...,
\eea 
where we include the contribution from the quark condensations in the limit $H=0$ where all quarks are massless.
To proceed, we can take the gauge that $H$ is identified as a real scalar field, and also choose the field basis where
the Yukawa couplings $y_\psi$ are real, positive and diagonal\footnote{The flavor changing weak interactions mediated by the W-boson in this field basis can be safely ignored.}. To avoid unnecessary complication due to nonzero tadpoles
of the fields other than  $\phi$, we then focus on the field configuration with
\bea
\tilde \pi_\psi = \frac{\pi}{2}\eea
where $\tilde\pi_\psi$ denotes the phase of the quark condensation for $H=0$, i.e.
\bea
\langle \bar \psi_L\psi_R \rangle_{H=0} =\tilde \Lambda^3 e^{i\tilde \pi_\psi} \eea
for $\tilde \Lambda$ which corresponds to  the QCD condensation scale for $H=0$, which is about an half of the QCD scale for $H=v$. 
For such field configuration, one immediately finds
\bea
\partial_H V(H=0, \tilde \pi_{\psi}=\pi/2) = \partial_{\tilde\pi_\psi}V(H=0, \tilde \pi_{\psi}=\pi/2)=0,\eea
which results in \cite{Denef:2018etk}:
\bea 
M_{\rm Pl}\frac{|\nabla V|}{V} \,=\, M_{\rm Pl}\frac{|V_{\rm eff}^\prime + \left({\lambda'}/{\lambda} + {4 v'}/{v}\right)\lambda v^4|} 
{V_{\rm eff}(\phi) + \lambda(\phi)v(\phi)^4}
 \,=\, \left|M_{\rm Pl}\left(\frac{\lambda'}{\lambda} + \frac{4 v'}{v}\right) + {\cal O}(10^{-55})\right| \,\gtrsim \,c.
\eea

Translating the above result to an observational bound on $c$ is more complicated and model-dependent than the case of the pion extremum.
Yet, with the matching condition  (\ref{eq:d_g})  on $d_g$ including the relevant radiative corrections and also 
the tree level matching condition  
$d_q/M_{\rm Pl} = m_q^\prime/m_q ={y_q^\prime}/{y_q}+{v^\prime}/{v}$,
we can estimate the maximal value of $c$ compatible with the observational constraints on the low energy quintessence couplings.
Given the observational bounds on $d_g$ and $d_q$, the maximal value of $c$ can be achieved 
when $|\lambda^\prime/\lambda|\gg |v^\prime/v|, |y_q^\prime/y_q|$ and $\lambda^\prime/\lambda$ saturates the bound on $d_g$ through the two-loop contribution represented by the last term of (\ref{eq:d_g}). In fact, the model discussed  in \cite{Denef:2018etk} corresponds to such case as it assumes $v^\prime = y_q^\prime =0$ with $\lambda^\prime\neq 0$.
Inserting all the involved numerical factors, we find that the corresponding bound on $c$ is given by 
\bea 
c \,\leq\, {\rm Max}\left[ 1.5 \times 10^4 d_g,\, 10^{-55} \right]\,\lesssim\,  4.4 \times 10^{-2}
\eea 
which is significantly weaker than the bound (\ref{eq:bound_gen}) from the pion extremum.
Note that for a metrical quintessence, we have
$|\lambda^\prime/\lambda|\ll |v^\prime/v|$ and the bound on $c$ from the Higgs extremum is same as the one from the pion extremum, i.e. $c <  1.4\times 10^{-2}$.



\section{\label{tadpole} Effects of the tadpole or runaway of additional scalar fields} 

In the previous section,  we discussed the implications of the pion or Higgs extremum for the dS swampland conjecture 
within an effective theory while assuming that other scalar degrees of  freedom can be integrated out in such a way that the resulting effective theory can describe well the relevant low energy physics over the entire field range of the pion field, i.e. $\pi_0/f_\pi \in [0,2\pi]$, and also over  the Higgs field range
$\Delta H \sim v$. Here we examine possible effects of the tadpole or runaway behavior of the integrated scalar fields, which can be induced at the pion or Higgs dS extremum. As it provides the most stringent bound on $c$, we will focus on the
case of the pion extremum. As we will see, the bounds on $c$ obtained in Sec. \ref{ds_extrema} can not be significantly relaxed by additional scalar fields when the observational constraints  are properly taken into account.

Let $\Phi$ denote a generic scalar field which is integrated out in the effective potential (\ref{eq:quint_pion}).
As the quintessence is the only rolling field in the present Universe, $\Phi$ should be properly stabilized at least 
when the pion field is at the vacuum with $\pi_0/f_\pi=0$. Then, one can always choose $\langle \Phi\rangle_{\pi_0=0}=0$ and  expand the full potential of $\phi, \pi_0$ and $\Phi$ as follows:
\bea\label{eq:quint_pion_Phi}
V(\phi, \pi_0, \Phi)= V_{\rm eff}(\phi)+ V_{\rm up}(\phi, \pi_0) + \frac{1}{2} m_{\Phi}^2(\phi) \Phi^2+ \left(
\frac{\Phi}{2\Lambda_\Phi(\phi)}  +\cdots\right)V_{\rm up}(\phi, \pi_0),
\eea 
where
\bea
&& V_{\rm up}(\phi, \pi_0) \equiv V(\phi, \pi_0, \Phi=0) -
V(\phi, \pi_0=0, \Phi=0) \nonumber \\
&\simeq & (m_u(\phi) +m_d(\phi))\Lambda^3(\phi) \left(1-\cos\frac{\pi_0}{f_\pi(\phi)}\right)= m_\pi^2(\phi)f_\pi^2(\phi) \left(1-\cos\frac{\pi_0}{f_\pi(\phi)}\right)
\eea
and the ellipsis denotes the terms higher order in $\Phi$. Obviously, here $m_\Phi$ is the mass of $\Phi$ when $\pi_0/f_\pi=0$, and $\Lambda_\Phi$ is a mass scale parametrizing the coupling of $\Phi$ to the pions or more generically to the low energy QCD sector.

If $m_\Phi\Lambda_\Phi > m_\pi f_\pi$,  $\Phi$ is stabilized with a small field shift even when the pion field is at $\pi_0/f_\pi=\pi$. The corresponding tadpole is determined by
\bea
\partial_{\pi_0} V(\phi, \pi_0/f_\pi =\pi, \Phi=\delta \Phi) =\partial_\Phi V(\phi, \pi_0/f_\pi =\pi, \Phi=\delta \Phi)
 =0,
\eea
yielding
\bea
\label{eq:tadpole}
\frac{\delta \Phi}{\Lambda_\Phi}\, \simeq \,\frac{m_\pi^2 f_\pi^2}{m_\Phi^2\Lambda_\Phi^2}. \eea
One may then apply the dS swampland conjecture for the shifted extremum point, which would result in
\bea \label{eq:swamp_at_pi}
&&M_{\rm Pl}\frac{|\nabla V(\phi, \pi_0/f_\pi=\pi, \Phi=\delta\Phi)|}{
 V(\phi, \pi_0/f_\pi=\pi, \Phi=\delta\Phi)}= M_{\rm Pl} \frac{|\partial_\phi V(\phi, \pi_0/f_\pi=\pi, \Phi=\delta\Phi)
 |}{V(\phi, \pi_0/f_\pi=\pi, \Phi=\delta\Phi)
 } 
 \nonumber \\
 &\simeq& M_{\rm Pl}\left|\partial_\phi \ln V_{\rm up}(\phi, \pi_0/f_\pi = \pi) 
- \frac{m_\pi^2f_\pi^2}{m_\Phi^2 \Lambda_\Phi^2} \partial_\phi \ln \left(\frac{V_{\rm up}(\phi, \pi_0/f_\pi = \pi)}{m_\Phi^2(\phi)\Lambda_\Phi^2(\phi)}\right)\right|
\nonumber\\
&=& M_{\rm Pl}\left|\left(\frac{m_u^\prime+m_d^\prime}{m_u+m_d}+3\frac{\Lambda^\prime}{\Lambda}\right)
-\frac{m_\pi^2  f_\pi^2 }{m_\Phi^2  \Lambda_\Phi^2}\left(\frac{m_u^\prime+m_d^\prime}{m_u+m_d}+3\frac{\Lambda^\prime}{\Lambda}
-\partial_\phi \ln (m_\Phi^2\Lambda_\Phi^2)\right)\right|\nonumber \\
&=&
\left|\,(d_{\tilde q}+ 3 d_g)\left(1-\frac{m_\pi^2  f_\pi^2 }{m_\Phi^2  \Lambda_\Phi^2}\right)
+\frac{m_\pi^2  f_\pi^2 }{m_\Phi^2  \Lambda_\Phi^2}
M_{\rm Pl}\partial_\phi \ln (m_\Phi^2\Lambda_\Phi^2)\,\right| \,\geq \, c.
\eea 
Note that the terms suppressed by $m_\pi^2 f_\pi^2/m_\Phi^2 \Lambda_\Phi^2$ correspond to the corrections to 
eq. (\ref{eq:pion_ds_bound}) in
Sec. \ref{ds_extrema}, which arise from the tadpole of $\Phi$ induced at $\pi_0/f_\pi =\pi$.

Let us apply the above results for the SM scalar degrees of freedom which are either elementary or composite.
Fist of all, for $\Phi$ being the pseudo-scalar mesons such as $K$ and $\eta$,
 P or CP symmetry assures that the linear coupling of $\Phi$ to $V_{\rm up}$ is highly suppressed, e.g. $\Lambda_\Phi \gg v$, and therefore 
$f_\pi^2m_\pi^2/m_\Phi^2\Lambda_\Phi^2\ll 10^{-5}$. 
For $\Phi$ being the SM Higgs boson, we have $\Lambda_\Phi\sim v$ and again
$m_\pi^2f_\pi/m_\Phi^2\Lambda_\Phi^2\sim m_\pi^2f_\pi^2/m_h^2 v^2 \ll 10^{-5}$.
 This assures the possible corrections due to the tadpole of the pseudo-scalar mesons and Higgs boson 
 are much smaller than the observational bound 
 on $d_g$ and $d_q$, and therefore can be safely ignored.
In fact, the only scalar degree of freedom of the SM which can have a non-negligible value of $f_\pi^2m_\pi^2/m_\Phi^2\Lambda_\Phi^2$
is the quark-antiquark composite scalar $\sigma$ which controls the size of the light quark condensation\footnote{In real QCD,
$\sigma$ has a too broad decay width, so there is no corresponding particle state. However, yet $\sigma$ can be relevant for the dS swampland conjecture as the potential energy varies as a function of $\sigma$.}:
\bea
\langle \bar q q\rangle \,\propto\, e^{-\sigma/\Lambda_\sigma}.\eea
for which 
\bea
m_\sigma \sim \Lambda_\sigma \sim \Lambda_{\rm QCD}.\eea
In this case, the corresponding suppression factor $m_\pi^2f_\pi^2/m_\sigma^2\Lambda_\sigma^2\sim m_q/\Lambda_{\rm QCD}$ is not small enough. However the accompanying factor which is given by
\bea
M_{\rm Pl}\partial_\phi \ln (m_\sigma^2 \Lambda_\sigma^2) = 4M_{\rm Pl}\frac{\Lambda^\prime_{\rm QCD}}{\Lambda_{\rm QCD}} = 4 d_g\eea
provides additional suppression, so that again the tadpole of $\sigma$ at the pion extremum does not alter  the result (\ref{eq:pion_ds_bound}).

Our discussion above suggests that the upper bound on $c$ can be relaxed if
there exists some scalar field $\Phi$ (other than those in the SM) with
$\frac{m_\pi^2  f_\pi^2 }{m_\Phi^2  \Lambda_\Phi^2}
M_{\rm Pl}\partial_\phi \ln (m_\Phi^2\Lambda_\Phi^2)\gg {\cal O}(10^{-5}-10^{-2})$ at $\pi_0/f_\pi=\pi$, i.e. a light scalar
which has a sizable tadpole or runaway behavior at the pion extremum.
In string theory, the most promising candidate for such scalar field is
a light modulus $\chi$. 
Even when $\chi$ has a sizable tadpole or runaway behavior at $\pi_0/f_\pi=\pi$, 
 it has to be stabilized at certain vacuum value $\langle\chi\rangle$
with a mass $m_\chi > H_0\sim 10^{-33}\, {\rm eV}$ at $\pi_0/f_\pi=0$. Again one can choose a field basis for which
$\langle \chi\rangle =0$ at $\pi_0/f_\pi=0$, and then
the potential  can be expanded as 
\bea\label{eq:quint_pion_chi}
V(\phi, \pi_0, \chi)= V_{\rm eff}(\phi)+ V_{\rm up}(\phi, \pi_0) + \frac{1}{2} m_{\chi}^2(\phi) \chi^2+ \left(
c_\chi\frac{\chi}{2M_{\rm Pl}}  +\cdots\right)V_{\rm up}(\phi, \pi_0),
\eea 
where we introduce a dimensionless parameter $c_\chi$ to parametrize the coupling of $\chi$ to the low energy QCD sector.
Here the coupling $c_\chi$ should be the low energy consequence of the modulus couplings to the QCD sector, which can be
parametrized as\bea
{\cal L}_\chi=
- \tilde d_g  \frac{\chi}{M_{\rm Pl}} \frac{\beta_s}{2g_s }G^{i\mu\nu}G^i_{\mu\nu}
- \sum_{q=u,d,s}(\tilde d_q +\gamma_m \tilde d_g) m_q\frac{\chi}{M_{\rm Pl}} \bar q q,
\eea
and then
\bea
c_\chi = \frac{m_u \tilde d_u + m_d\tilde d_d}{m_u+m_d} + 3 \tilde d_g\equiv \tilde d_{\tilde q} + 3\tilde d_g .
\eea

There are some range of $m_\chi$ for which the modulus $\chi$ is obviously in conflict with the observational constraints or not useful for relaxing the bound on $c$. For instance,
for a relatively massive $\chi$ with
\bea
m_\chi \gtrsim  1.5\times 10^{-9} c_\chi\, {\rm eV},\eea
one can apply (\ref{eq:swamp_at_pi}) with $m_\Phi=m_\chi$ and $\Lambda_\Phi=M_{\rm Pl}/c_\chi$ to ensure that
the modulus tadpole $\delta\chi$ is small enough to keep the bound (\ref{bound_pion_ex}) unaffected.
Also, the following mass regions are excluded by the blackhole superradiance \cite{Arvanitaki:2014wva}:
 \bea 
 \label{bh_sr}
 5 \times 10^{-13}\,{\rm eV} \lesssim m_\chi  \lesssim 2\times10^{-11}\, {\rm eV}, \nonumber \\
 10^{-17}\,{\rm eV} \lesssim m_\chi  \lesssim  6\times10^{-17}\,{\rm eV},\nonumber \\
 8 \times 10^{-19}\,{\rm eV} \lesssim m_\chi  \lesssim  10^{-18}\,{\rm eV}.
\eea

If $\chi$ is light enough, e.g. 
\bea
m_\chi < 5\times 10^{-12}c_\chi^{1/2}\, {\rm eV},\eea
the resulting  modulus tadpole $\delta \chi$ is of ${\cal O}(M_{\rm Pl})$, so 
 a nearby  stationary point is not guaranteed to exist. In such case, one can apply the dS swampland conjecture to the  field configuration with $\pi_0/f_\pi=\pi$ and $\chi=0$, rather than the  configuration with $\pi_0/f_\pi=\pi$ and $\chi=\delta\chi$.
This then leads to a new upper bound on $c$ as
\bea
\label{new_bound}
&& M_{\rm Pl}\frac{|\nabla V(\phi, \pi_0/f_\pi= \pi, \chi=0)|}{V(\phi, \pi_0/f_\pi= \pi, \chi=0)}\, =\, M_{\rm Pl}\frac{\sqrt{\left({\partial_\phi V}\right)^2 
	+ \left({\partial_\chi V}\right)^2}}{V}  \nonumber \\
&& \hskip 1cm \,\simeq\, \sqrt{ 
(d_{\tilde q} + 3d_g)^2
+ c_\chi^2}\, \geq\,  c.
\eea 
For $m_\chi < 10^{-16}$ eV, the modulus couplings $\tilde d_g$ and $\tilde d_q$ are constrained as the quintessence couplings
$d_q$ and $d_g$
by both the non-observation of the violation of the equivalence principle \cite{Touboul:2017grn} and the gravitational time delay in the solar system
\cite{Bertotti:2003rm,Hohmann:2013rba, Scharer:2014kya}. Then the above new bound is essentially  equivalent to the bound (\ref{bound_pion_ex}), which means that a ultralight modulus with
$m_\chi< 10^{-16}$ eV is not useful for relaxing the bound (\ref{bound_pion_ex}).
If $(\tilde d_g- \tilde d_q)/(\tilde d_g+\tilde d_q)={\cal O}(1)$, so that the modulus couplings violate the equivalence principle, one can use the corresponding bounds on $\tilde d_g$ and $\tilde d_q$ to get the following bound on $c$ from (\ref{new_bound}): 
 \bea
 c \,< \, 2\times10^{-5} \quad{\rm for}\quad m_\chi< 10^{-13} \, {\rm eV}.  \eea
This means that a modulus with $m_\chi< 10^{-13}$ eV and $(\tilde d_g- \tilde d_q)/(\tilde d_g+\tilde d_q)={\cal O}(1)$
is again not useful for relaxing the bound (\ref{bound_pion_ex}).

The light modulus $\chi$ which may have a sizable tadpole or runaway behavior at the pion extremum is  constrained also  by the cosmological modulus mass density
generated by the modulus misalignment $\Delta \chi$ in the early Universe which is induced by the  coupling
$c_\chi$. Note that a nonzero $c_\chi$ means that $\chi$ couples to the gluons and/or the light quarks, so the thermal free energy of gluons and light quarks in the early Universe depends on $\chi$. Such modulus-dependent free energy induces a modulus misalignment, which eventually produces the modulus dark matter. To examine this issue, let us consider
the finite temperature effective potential before the QCD phase transition, which is given by
\bea
V(\chi,T)= -\frac{\pi^2}{90}g_* (T) T^4 + \frac{2 T^2}{3}   \left(3+\frac{N_f}{2}\right) \frac{g_s^2(\chi) T^2}{6} + \sum_{m_q <T} \frac{T^2}{4} \left( m_q^2(\chi) + \frac{ g_s^2(\chi) T^2}{6}  \right  ),
\eea
where $g_*$ is the number of degrees of freedom contributing to the free energy and $N_f$ is the number of quark flavors lighter than $T$.
This potential provides a slope to the modulus $\chi$ as
\bea
\frac{\partial V}{\partial\chi}= \frac{2 \alpha_s^2 T^4}{9}  \left( 3 + \frac{7}{8} N_f \right) \left( b_3 \frac{\tilde{d}_g}{M_{\rm Pl}} -\sum_{\Lambda_{\rm QCD} < m_q < T} \frac{2}{3} \partial_\chi \ln (y_q v) \right)
+ \sum_{m_q<T} \frac{\tilde{d}_q m_q^2 T^2}{2 M_{\rm Pl}} ,
\eea
which induces a modulus misalignment
\bea\label{eq:misalignment}
\frac{\Delta \chi}{M_{\rm Pl}} \simeq  \frac{\partial_\chi V}{M_{\rm Pl} H^2} .
\eea 
We estimate such modulus misalignment at $T=T_\chi \simeq \sqrt{m_\chi M_{\rm Pl}}$
 and find
 \bea
 \frac{\Delta \chi}{M_{\rm Pl}} &\gtrsim & 0.5 \frac{\left( 3+\frac{7}{8}N_f \right) \left( 11-\frac{2}{3}N_f \right)}{g_* (T_\chi)} \alpha_s^2 ( T_\chi )  c_\chi + \sum_{m_q > T_\chi} \frac{0.4}{\sqrt{g_* (m_q)}}c_\chi    \quad \mbox{for}\quad m_\chi > 10^{-11}{\rm eV} ,\nonumber \\
 \frac{\Delta \chi}{M_{\rm Pl}} &\gtrsim & 0.4 \, c_\chi \quad \mbox{for}\quad m_\chi \ll  10^{-11}{\rm eV},
 \eea
 where for simplicity $\tilde{d}_q$ and $\tilde{d}_g$ are assumed to have a similar value.
Requiring that the resulting modulus mass density does not exceed the observed dark matter mass density, i.e.
\bea
\Omega_\chi h^2 \, <\, 0.12,\eea
we obtain the following bounds on the modulus coupling:
\bea
c_\chi  &\lesssim & \frac{2.2 \times 10^{-6}}{  \alpha_s^2 ( T_\chi ) g_*^{-1}(T_\chi) + \sum_{m_q > T_\chi}0.015 g_*^{-1/2}(m_q)}  \left( \frac{10^{-12} {\rm eV}}{m_\chi} \right)^{\frac{1}{4}}   \quad \mbox{for}\quad m_\chi > 10^{-11}{\rm eV},\nonumber \\
c_\chi &\lesssim & 1.4 \times 10^{-4} \left( \frac{10^{-12} {\rm eV}}{m_\chi} \right)^{\frac{1}{4}} \quad \mbox{for}\quad m_\chi \ll  10^{-11}{\rm eV}.\eea

 \begin{figure}[t]
\includegraphics[width=0.8\textwidth]{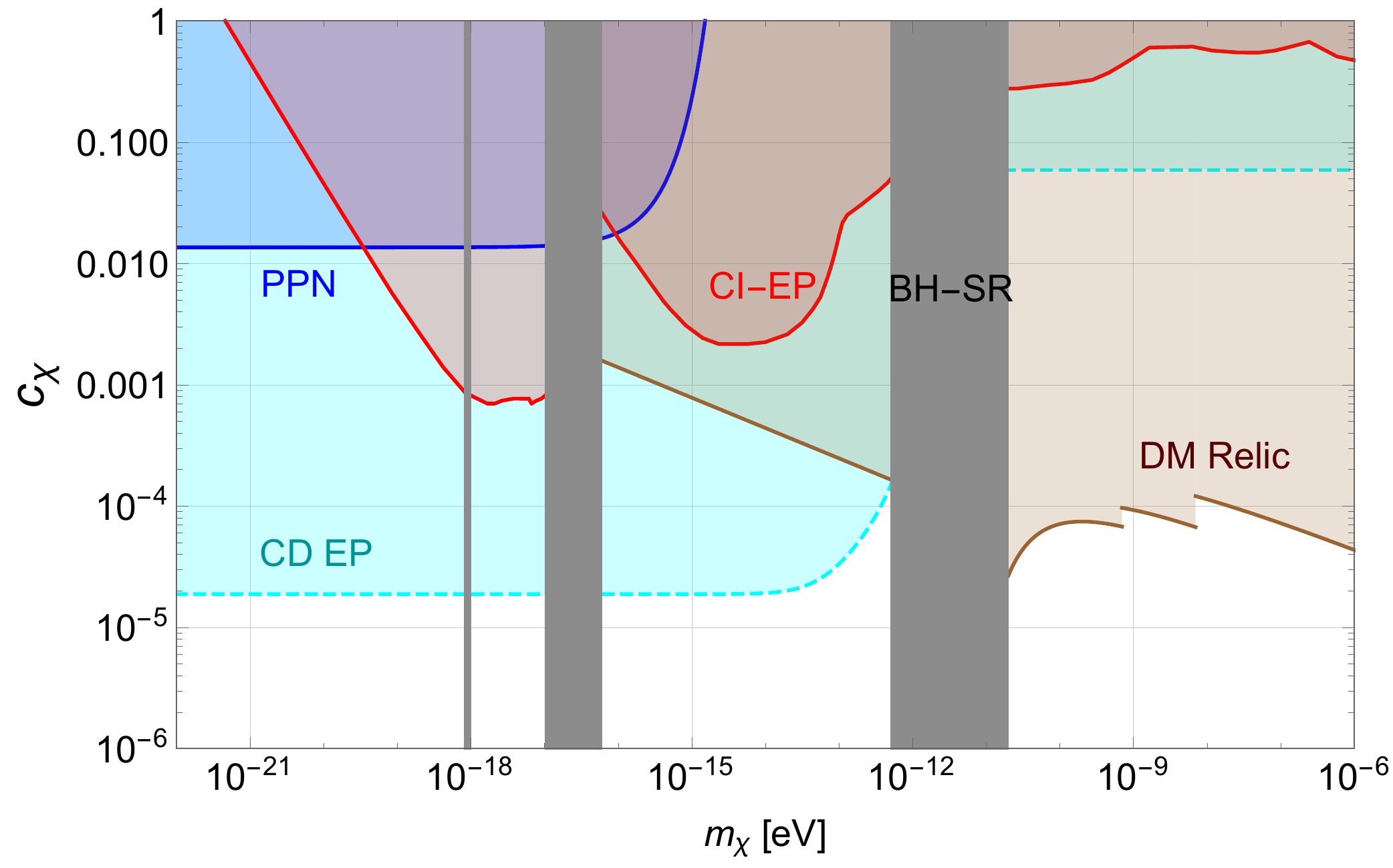}
\caption{Observational upper bound on the modulus coupling $c_\chi=\tilde{d}_{\tilde{q}}+3\tilde{d}_g$ as a function of the modulus mass $m_\chi$. The gray regions are excluded by the blackhole superradiance (BH-SR) \cite{Arvanitaki:2014wva}. The cyan region and the red region are excluded, respectively, by the composition-dependent \cite{Berge:2017ovy,Smith:1999cr} and composition-independent \cite{Fischbach:1996eq,Arvanitaki:2014faa}  equivalence principle (CD-EP and CI-EP) tests. Constraints on the parametrized post-Newtonian (PPN) parameters \cite{Bertotti:2003rm,Hohmann:2013rba, Scharer:2014kya} exclude the blue region, which applies not only for a generic modulus, but also for a metrical modulus respecting the equivalence principle. The dark matter relic abundance constrains the modulus misalignment, excluding the brown region.}
\label{ModuliCoupling}
\end{figure}

 In Fig.~\ref{ModuliCoupling}, we summarize the available constraints on the modulus coupling $c_\chi$ as a function of $m_\chi$.
 The blackhole superradiance (BH-SR)~\cite{Arvanitaki:2014wva} excludes the modulus mass range (\ref{bh_sr}) which corresponds to the gray region
 in Fig.~\ref{ModuliCoupling}.
 The cyan region bounded by a dotted line for $m_\chi<10^{-12}$~eV is excluded by the MICROSCOPE test  of the composition-dependent
  equivalence principle  (CD-EP)~\cite{Berge:2017ovy}, while the region above $10^{-12}$~eV
  is excluded by the short range test of the CD-EP~\cite{Smith:1999cr}. 
 If $\chi$ respects the equivalence principle,
 e.g. a metrical modulus $\chi$ with $\tilde{d}_q = \tilde{d}_g$, the CD-EP bounds do not apply anymore. Yet, for certain range of $m_\chi$, such metrical modulus can result in an observable deviation of the gravitational potential from $1/r$, which is bounded by the composition-independent equivalence principle (CI-EP) test.  The red region of Fig.~\ref{ModuliCoupling} is excluded by such experiments testing the CI-EP 
 \cite{Fischbach:1996eq,Arvanitaki:2014faa}. The blue region of 
Fig.~\ref{ModuliCoupling} 
  is excluded by the measurement of the gravitational time delay to photons from the Cassini spacecraft in the parametrized post-Newtonian (PPN) formalism \cite{Bertotti:2003rm,Hohmann:2013rba, Scharer:2014kya}, which applies not only for generic modulus, but also for a metrical modulus which respects the equivalence principle. Finally, the brown region is excluded by
 the bound on the relic modulus dark matter produced by the modulus misalignment induced by the coupling $c_\chi$.
  
 \begin{figure}[t]
\includegraphics[width=0.8\textwidth]{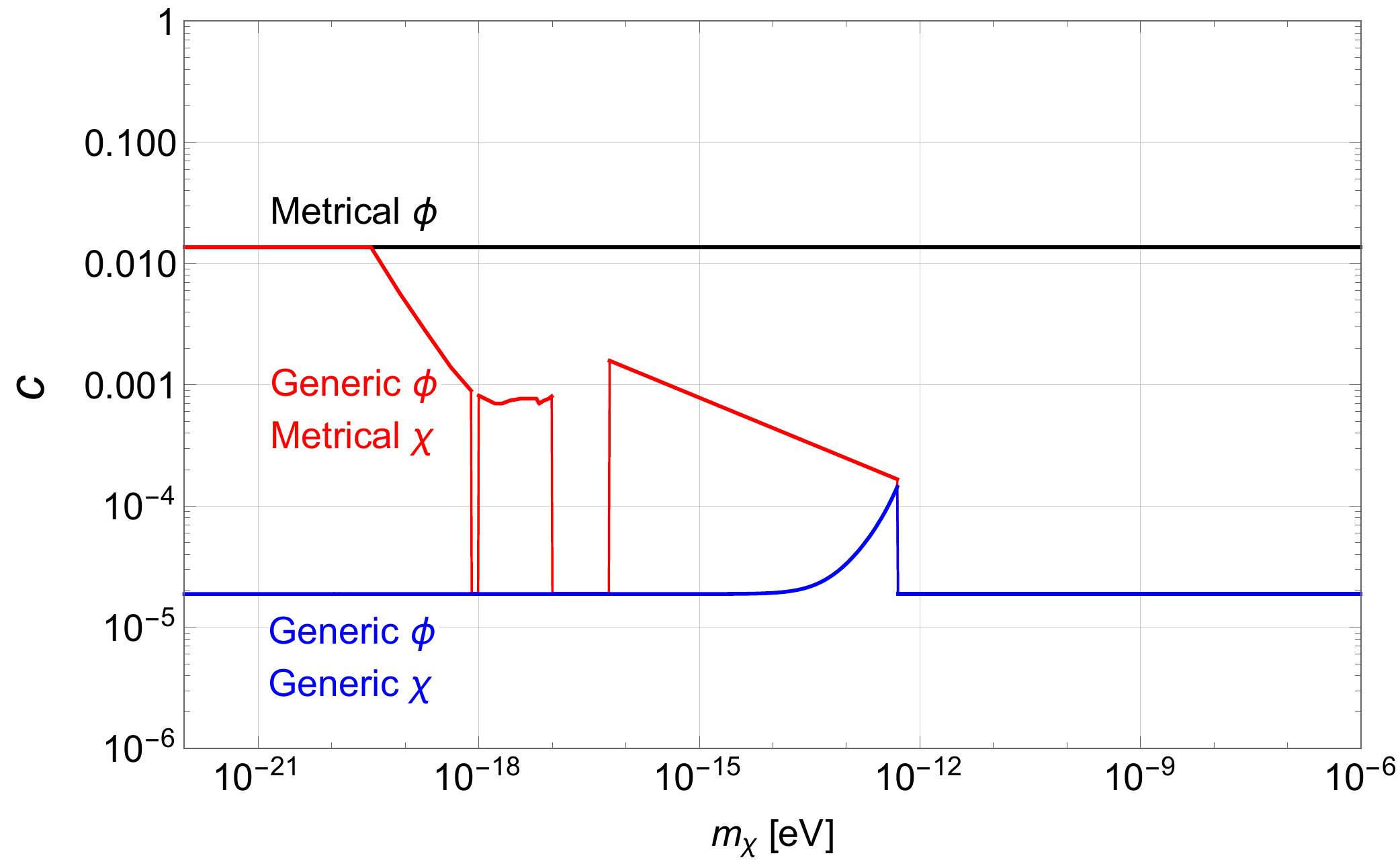}
\caption{The refined upper bound on the parameter $c$ as a function of the mass $m_\chi$ of a modulus-like scalar field $\chi$ introduced to relax the original bounds (\ref{eq:bound_gen}) and (\ref{eq:bound_met}). This shows that the original bounds are not significantly relaxed by additional light scalar once the observational constraints on such light scalar are properly taken into account.}
\label{Ccases}
\end{figure}

Since its coupling $c_\chi$ is severely constrained as above, the modulus-like scalar $\chi$ can not significantly relax the bounds  (\ref{eq:bound_gen}) and (\ref{eq:bound_met})  which were obtained in the effective theory  where $\chi$ is integrated out.  In Fig.~\ref{Ccases}, we depict the refined bound on $c$ taking into account the effects of  $\chi$
for three different cases. The blue line is the bound for the case that both the quintessence $\phi$ and the additional modulus $\chi$ have generic couplings violating the equivalence principle, while the black (red) line corresponds to the case that $\phi$ ($\chi$) respects the equivalence principle. 
The results of Fig.~\ref{Ccases} can be extrapolated to $m_\chi\ll 10^{-21}$ eV in a straightforward manner. On the other hand, for
$m_\chi> 10^{-6}$ eV, one can use (\ref{eq:tadpole}) and (\ref{eq:swamp_at_pi}) to assure that the bounds  (\ref{eq:bound_gen}) and (\ref{eq:bound_met}) can not be relaxed by introducing $\chi$.

With the above results, let us now regard  (\ref{eq:bound_gen}) as the bound on $c$
when both $\phi$ and $\chi$ have generic couplings to violate the equivalence principle, while
(\ref{eq:bound_gen}) corresponds to the bound when any of $\phi$ and $\chi$ respects the equivalence principle. Then our results imply
that  those bounds on $c$ 
can not be significantly relaxed by additional modulus-like light scalar field once the observational constraints on such scalar field are properly taken into account\footnote{Our result suggests that the stronger bound (\ref{eq:bound_gen}) can be relaxed by about one order of magnitude by a modulus $\chi$ with $m_\chi={\cal O}(10^{-13})$ eV. However this minor point does not change the main message of our results.}. 


\section{\label{conclusion} conclusion}
 
In this paper
 we examined the implications of the pion or Higgs dS extrema  for the dS swampland conjecture, while focusing
 on the possible  (model-independent or model-dependent) 
 bound on the parameter $c$. 
Applying the dS swampland conjecture to the pion  extremum at $\cos(\pi_0/f_\pi)=-1$, we could derive a model-independent upper bound on $c$ given in terms of the low energy quintessence couplings. 

If the quintessence couplings take a rather specific form to respect the equivalence principle, which would be the case when the quintessence couples to the SM sector {\it} only through the trace of the energy momentum tensor, $c$ is bounded essentially by the observational bound on the deviation from the general relativity by the quintessence-mediated force in relativistic limit, yielding 
$c< 1.4\times 10^{-2}$. However, for generic quintessence whose couplings violate the equivalence principle, the parameter $c$ is more strongly bounded as $c< 2\times 10^{-5}$ by the non-observation of the violation of the equivalence principle in non-relativistic limit. 
These bounds on $c$ are rather robust as (i) they are obtained within the framework of the most general quintessence potential and couplings and (ii) they can not be significantly relaxed by an additional light scalar field which may have a nonzero tadpole or runaway behavior at the pion dS extremum, if the observational constraints on such light scalar field are properly taken into account.
One can do a similar analysis for the Higgs extremum at $H=0$, but the resulting bound $c\lesssim 4.4 \times 10^{-2}$ is weaker than those from the pion extremum as the quintessence couplings to the Higgs sector is more weakly constrained than those to the low energy QCD sector.

Our bounds on $c$ from the pion extremum appear to have a significant tension with the dS swampland conjecture (\ref{conjecture}) which assumes $c={\cal O}(1)$. Yet the conjecture (\ref{conjecture}) can be modified or refined, for instance as in  \cite{Dvali:2018fqu},\cite{Andriot:2018wzk} and \cite{Garg:2018reu}, which would allow us to avoid the bounds from the Higgs and pion extrema. If it is the right direction to pursue, our results provide additional motivation for such refinement of the conjecture.

\begin{acknowledgments}
This work was supported by IBS under the project code IBS-R018-D1. K.C. thanks A. Hebecker for useful discussions. C.S.S. thanks Deog Ki Hong for valuable discussions during the CERN-Korea TH Institute, and also the CERN Theory Group for the hospitality. 
\end{acknowledgments}


\begin{thebibliography}{00}
	
	\bibliographystyle{JHEP}
	
	
	
	\bibitem{Obied:2018sgi} 
	G.~Obied, H.~Ooguri, L.~Spodyneiko and C.~Vafa,
	arXiv:1806.08362 [hep-th].
	
	\bibitem{Agrawal:2018own} 
	P.~Agrawal, G.~Obied, P.~J.~Steinhardt and C.~Vafa,
	Phys.\ Lett.\ B {\bf 784}, 271 (2018)
	doi:10.1016/j.physletb.2018.07.040
	[arXiv:1806.09718 [hep-th]].
	
\bibitem{Ratra:1987rm} 
  B.~Ratra and P.~J.~E.~Peebles,
  Phys.\ Rev.\ D {\bf 37}, 3406 (1988).
  doi:10.1103/PhysRevD.37.3406
  
\bibitem{Wetterich:1987fm} 
  C.~Wetterich,
  Nucl.\ Phys.\ B {\bf 302}, 668 (1988)
  doi:10.1016/0550-3213(88)90193-9
  [arXiv:1711.03844 [hep-th]].
	
\bibitem{Zlatev:1998tr} 
  I.~Zlatev, L.~M.~Wang and P.~J.~Steinhardt,
  Phys.\ Rev.\ Lett.\  {\bf 82}, 896 (1999)
  doi:10.1103/PhysRevLett.82.896
  [astro-ph/9807002].
	
	\bibitem{Denef:2018etk} 
	F.~Denef, A.~Hebecker and T.~Wrase,
	arXiv:1807.06581 [hep-th].
	
	\bibitem{Dvali:2018fqu} 
 	 G.~Dvali and C.~Gomez,
 	 arXiv:1806.10877 [hep-th].
	
	\bibitem{Andriot:2018wzk} 
  	D.~Andriot,
 	 arXiv:1806.10999 [hep-th].
  
  
  
  
  
  \bibitem{Garg:2018reu} 
  S.~K.~Garg and C.~Krishnan,
  arXiv:1807.05193 [hep-th].
  
  
  
  
  
  
  
 
\bibitem{ds_swc} 

 
  S.~Banerjee, U.~Danielsson, G.~Dibitetto, S.~Giri and M.~Schillo,
  arXiv:1807.01570 [hep-th];
  L.~Aalsma, M.~Tournoy, J.~P.~Van Der Schaar and B.~Vercnocke,
  arXiv:1807.03303 [hep-th];
  A.~Achúcarro and G.~A.~Palma,
  arXiv:1807.04390 [hep-th];
  J.~L.~Lehners,
  arXiv:1807.05240 [hep-th];
  A.~Kehagias and A.~Riotto,
  arXiv:1807.05445 [hep-th];
  M.~Dias, J.~Frazer, A.~Retolaza and A.~Westphal,
  arXiv:1807.06579 [hep-th];
  E.~Ó.~Colgáin, M.~H.~P.~M.~Van Putten and H.~Yavartanoo,
  arXiv:1807.07451 [hep-th];
  C.~Roupec and T.~Wrase,
  arXiv:1807.09538 [hep-th];
  D.~Andriot,
  arXiv:1807.09698 [hep-th];
  H.~Matsui and F.~Takahashi,
  arXiv:1807.11938 [hep-th].
  I.~Ben-Dayan,
  arXiv:1808.01615 [hep-th];
  O.~Loaiza-Brito and O.~Loaiza-Brito,
  arXiv:1808.03397 [hep-th];
  J.~P.~Conlon,
  arXiv:1808.05040 [hep-th];
  W.~H.~Kinney, S.~Vagnozzi and L.~Visinelli,
  arXiv:1808.06424 [astro-ph.CO];
  K.~Dasgupta, M.~Emelin, E.~McDonough and R.~Tatar,
  arXiv:1808.07498 [hep-th];
  S.~Kachru and S.~Trivedi,
  arXiv:1808.08971 [hep-th];
L.~Heisenberg, M.~Bartelmann, R.~Brandenberger and A.~Refregier,
arXiv:1809.00154 [astro-ph.CO]; 
H.~Murayama, M.~Yamazaki and T.~T.~Yanagida,
arXiv:1809.00478 [hep-th];
 M.~C.~D.~Marsh,
 arXiv:1809.00726 [hep-th].


	\bibitem{Choi:1999xn} 
  	K.~Choi,
  	Phys.\ Rev.\ D {\bf 62}, 043509 (2000)
  	doi:10.1103/PhysRevD.62.043509
  	[hep-ph/9902292].
	
	\bibitem{Choi:1999wv} 
	  K.~Choi,
	  hep-ph/9912218.
	
	\bibitem{Chiang:2018jdg} 
  	C.~I.~Chiang and H.~Murayama,
  	arXiv:1808.02279 [hep-th].
  
	\bibitem{Cicoli:2018kdo} 
	  M.~Cicoli, S.~de Alwis, A.~Maharana, F.~Muia and F.~Quevedo,
	  arXiv:1808.08967 [hep-th].
  
	\bibitem{Akrami:2018ylq} 
	 Y.~Akrami, R.~Kallosh, A.~Linde and V.~Vardanyan,
	 arXiv:1808.09440 [hep-th].
  
\bibitem{Bertotti:2003rm} 
  B.~Bertotti, L.~Iess and P.~Tortora,
  Nature {\bf 425}, 374 (2003).
  doi:10.1038/nature01997
	
	\bibitem{Godun:2014naa} 
	R.~M.~Godun {\it et al.},
	Phys.\ Rev.\ Lett.\  {\bf 113}, no. 21, 210801 (2014)
	doi:10.1103/PhysRevLett.113.210801
	[arXiv:1407.0164 [physics.atom-ph]].
	
	
	\bibitem{Touboul:2017grn} 
	P.~Touboul {\it et al.},
	Phys.\ Rev.\ Lett.\  {\bf 119}, no. 23, 231101 (2017)
	doi:10.1103/PhysRevLett.119.231101
	[arXiv:1712.01176 [astro-ph.IM]].
	
\bibitem{Berge:2017ovy} 
  J.~Bergé, P.~Brax, G.~Métris, M.~Pernot-Borràs, P.~Touboul and J.~P.~Uzan,
  Phys.\ Rev.\ Lett.\  {\bf 120}, no. 14, 141101 (2018)
  doi:10.1103/PhysRevLett.120.141101
  [arXiv:1712.00483 [gr-qc]].
	
	
	
	
	\bibitem{Uzan:2010pm} 
	  J.~P.~Uzan,
	  Living Rev.\ Rel.\  {\bf 14}, 2 (2011)
	  doi:10.12942/lrr-2011-2
	  [arXiv:1009.5514 [astro-ph.CO]].
	
	\bibitem{Heisenberg:2018yae} 
	  L.~Heisenberg, M.~Bartelmann, R.~Brandenberger and A.~Refregier,
	  arXiv:1808.02877 [astro-ph.CO].
  
\bibitem{Damour:2010rm} 
  T.~Damour and J.~F.~Donoghue,
  Class.\ Quant.\ Grav.\  {\bf 27}, 202001 (2010)
  doi:10.1088/0264-9381/27/20/202001
  [arXiv:1007.2790 [gr-qc]].
  
\bibitem{Damour:2010rp} 
  T.~Damour and J.~F.~Donoghue,
  Phys.\ Rev.\ D {\bf 82}, 084033 (2010)
  doi:10.1103/PhysRevD.82.084033
  [arXiv:1007.2792 [gr-qc]].
  
  
	
	\bibitem{Hohmann:2013rba} 
	 M.~Hohmann, L.~Jarv, P.~Kuusk and E.~Randla,
	 Phys.\ Rev.\ D {\bf 88}, no. 8, 084054 (2013)
	 Erratum: [Phys.\ Rev.\ D {\bf 89}, no. 6, 069901 (2014)]
	 doi:10.1103/PhysRevD.89.069901, 10.1103/PhysRevD.88.084054
	 [arXiv:1309.0031 [gr-qc]].
	
	\bibitem{Scharer:2014kya} 
	 A.~Schärer, R.~Angélil, R.~Bondarescu, P.~Jetzer and A.~Lundgren,
	 Phys.\ Rev.\ D {\bf 90}, no. 12, 123005 (2014)
	 doi:10.1103/PhysRevD.90.123005
	 [arXiv:1410.7914 [gr-qc]].
	
	\bibitem{Arvanitaki:2014wva} 
	  A.~Arvanitaki, M.~Baryakhtar and X.~Huang,
	  Phys.\ Rev.\ D {\bf 91}, no. 8, 084011 (2015)
	  doi:10.1103/PhysRevD.91.084011
	  [arXiv:1411.2263 [hep-ph]].
	
\bibitem{Smith:1999cr} 
  G.~L.~Smith, C.~D.~Hoyle, J.~H.~Gundlach, E.~G.~Adelberger, B.~R.~Heckel and H.~E.~Swanson,
  Phys.\ Rev.\ D {\bf 61}, 022001 (2000).
  doi:10.1103/PhysRevD.61.022001
	
\bibitem{Fischbach:1996eq} 
  E.~Fischbach and C.~Talmadge,
  hep-ph/9606249.
  
\bibitem{Arvanitaki:2014faa} 
  A.~Arvanitaki, J.~Huang and K.~Van Tilburg,
  Phys.\ Rev.\ D {\bf 91}, no. 1, 015015 (2015)
  doi:10.1103/PhysRevD.91.015015
  [arXiv:1405.2925 [hep-ph]].
	
	
\end{thebibliography}
\end{document}